\RequirePackage{rotating}
\documentclass[fleqn,usenatbib]{mnras}

\usepackage[T1]{fontenc}
\usepackage{ae,aecompl}

\usepackage{newtxtext}
\usepackage[varvw]{newtxmath}

\usepackage{graphicx}	
\usepackage{amsmath}	
\usepackage[british]{babel}
\usepackage[]{units}
\usepackage{caption}
\usepackage{booktabs}
\usepackage{multirow}
\usepackage{rotating}
\usepackage[para,referable]{threeparttablex}
\usepackage{xcolor}
\usepackage{microtype}
\hypersetup{pdfauthor={Errani et al.},
            pdftitle={},
            pdfkeywords={},
            bookmarksnumbered=true}

\newcommand{\rmx}{r_\mathrm{mx}}	%
\newcommand{\rhomx}{\rho_\mathrm{mx}}	%
\newcommand{\Vmx}{V_\mathrm{mx}}	%
\newcommand{\Mmx}{M_\mathrm{mx}}	%
\newcommand{\Tmx}{T_\mathrm{mx}}	%
\newcommand{\Tgeo}{T_\mathrm{geo}}	%
\newcommand{\kms}{\mathrm{km\,s^{-1}}}		%
\newcommand{\diff}{\mathrm{d}}
\newcommand{\Msol}{\mathrm{M_{\odot}}}

\newcommand{\rc}{r_\mathrm{c}}
\newcommand{\rs}{r_\mathrm{s}}
\newcommand{\Vc}{V_\mathrm{c}}

\newcommand{\kpc}{\mathrm{kpc}}
\newcommand{\pc}{\mathrm{pc}}
\newcommand{\Gyr}{\mathrm{Gyr}}
\newcommand{\rperi}{r_\mathrm{peri}}
\newcommand{\Tperi}{T_\mathrm{peri}}
\newcommand{\Torb}{T_\mathrm{orb}}
\newcommand{\rapo}{r_\mathrm{apo}}
\newcommand{\mxzero}{_\mathrm{mx0}}
\newcommand{\rt}{{\,:\,}}

\newcommand{\cm}{\mathrm{cm}}
\newcommand{\g}{\mathrm{g}}

\title[Dark matter cores and tidal survival]{Dark matter halo cores and the tidal survival of Milky Way satellites}

\author[]{Rapha\"el Errani${}^{1}$\thanks{errani@unistra.fr},
Julio F. Navarro${}^2$, 
Jorge Pe\~narrubia${}^3$,
Benoit Famaey${}^{1}$,
Rodrigo Ibata${}^{1}$
\\
$^1$ Universit\'e de Strasbourg, CNRS, Observatoire Astronomique de Strasbourg, UMR 7550, F-67000 Strasbourg, France\\
$^2$ Department of Physics and Astronomy, University of Victoria, Victoria, BC V8P 5C2, Canada\\
$^3$ Institute for Astronomy, University of Edinburgh, Royal Observatory, Blackford Hill, Edinburgh EH9 3HJ, UK
}

\usepackage{datetime}
\date{Accepted 2022 November 23. Received 2022 November 21; in original form 2022 October 1}

\pubyear{2022}
\volume{}

\begin{document}

\label{firstpage}
\pagerange{\pageref{firstpage}--\pageref{lastpage}} \pubyear{2022}
\maketitle

\begin{abstract}
The cuspy central density profiles of cold dark matter (CDM) haloes make them highly resilient to disruption by tides. Self-interactions between dark matter particles, or the cycling of baryons, may result in the formation of a constant-density core that would make haloes more susceptible to tidal disruption. We use N-body simulations to study the evolution of Navarro-Frenk-White (NFW)-like ``cored'' subhaloes in the tidal field of a massive host, and identify the criteria and time-scales for full disruption. Our results imply that the survival of Milky Way satellites places constraints on the sizes of dark matter cores. We find that no subhaloes with cores larger than 1 per cent of their initial NFW scale radius can survive for a Hubble time on orbits with pericentres $\lesssim 10\,\kpc$. A satellite like Tucana~3, with pericentre $\sim 3.5\,\kpc$, must have a core size smaller than $\sim 2\,\pc$ to survive just three orbital periods on its current orbit. The core sizes expected in self-interacting dark matter (SIDM) models with a velocity-independent cross-section of $1\,\mathrm{cm}^{-2}\mathrm{g}^{-1}$ seem incompatible with ultrafaint satellites with small pericentric radii, such as Tuc~3, Seg~1, Seg~2, Ret~2, Tri~2, and Wil~1, as these should have fully disrupted if accreted on to the Milky Way $\gtrsim 10\,\Gyr$ ago. These results suggest that many satellites have vanishingly small core sizes, consistent with CDM cusps. The discovery of further Milky Way satellites on orbits with small pericentric radii would strengthen these conclusions and allow for stricter upper limits on the core sizes.
\end{abstract}


\defcitealias{EN21}{EN21}

\section{Introduction}
\label{SecIntro}

In the Lambda cold dark matter (LCDM) cosmology, dark matter structures are organized in a hierarchy of haloes and subhaloes spanning a wide range of masses. LCDM halo density profiles are ``universal'', in the sense that the profile shape does not depend on mass, size or redshift, and ``cuspy'', in the sense that their central densities formally diverge, approximately following a Navarro-Frenk-White (NFW) profile \citep[][]{Navarro1996a, Navarro1997}. 

The cuspy nature of haloes around galaxies has long been a matter of controversy, with a large body of work arguing that this basic LCDM prediction is inconsistent with the slowly rising rotation curves of some dwarf and low-surface brightness galaxies (\citealt{Flores1994}; \citealt{Moore1994}; \citealt{deBlok2001}; \citealt{Oh2011}; \citealt{ReadCores2017}; for a review, see \citealt{deBlok2010}), as well as with the mass profiles inferred from kinematic analyses of dwarf spheroidal (dSph) galaxies (e.g. \citealt{Gilmore2007}; for a review see \citealt{Boldrini2022}), in particular for the Fornax and Sculptor dSphs \citep{Walker2011, Amorisco2012, AmoriscoAgnelloEvans2013,Diakogiannis2017}.

These studies have suggested that the rotation curve or velocity dispersion data are best accommodated by dark matter haloes with a constant density ``core'', although this conclusion has been challenged by other studies. In particular, \citet{Oman2015} argue that dwarf galaxy rotation curves are diverse, and that only \emph{some} dwarfs seem to indicate the presence of cores whereas others are consistent with cusps \citep[see also][]{Ghari2019}. Further work has also hinted that the observed rotation curve diversity may be at least partially driven by non-circular motions in dwarf galaxy discs \citep{Valenzuela2007,Oman2019,Santos-Santos2020,Roper2022}. 
The presence of cores in dSphs has also been disputed, e.g., in the case of the Sculptor dSph, by the models of \citet{RichardsonFairbairn2014} and \citet{Strigari2017}. This controversy is perhaps not surprising, as dynamic models of the inner density structure of dSphs are complicated by degeneracies introduced by the availability, in most cases, of only line-of-sight velocities and projected positions for individual stars \citep{Strigari2007,Laporte2013,Diakogiannis2014a, Read2017,EPW18,Genina2018,Genina2020}.

Similarly, some studies have concluded that the dynamical friction time-scales of globular clusters in the Fornax dSph \citep{Read2006,Cole2012}, as well as in the ultrafaint dwarf Eridanus~2 \citep{Contenta2018}, favour dark matter cores at their centres, while others have argued that the existent data cannot be used to conclusively rule out cuspy profiles \citep{Angus2009, Meadows2020}.

In spite of this unsettled state of affairs, constant-density cores are of considerable theoretical interest. Baryonic feedback (e.g., supernova explosions), for example, may drive gas out of dwarf galaxy central regions, leading to fluctuations in the overall potential that may soften the dark matter cusp and lead to the formation of a core \citep[see e.g.][]{Navarro1996b,Mashchenko2008, Penarrubia2012, Pontzen2012, Onorbe2015, El-Zant2016, Read2016cores, Read2019, Orkney2021}. 
On the other hand, if dark matter has a finite and sizable cross section for self-interaction, the repeated collisions between particles could thermalize the inner halo region, resulting in an isothermal central core \citep[self-interacting dark matter or SIDM; see e.g.][]{Burkert2000, Spergel2000, Vogelsberger2012, Rocha2013, Kahlhoefer2019,Kaplinghat2020}.
Cores are therefore potential probes of either the baryon cycling during galaxy formation, or of the nature of dark matter, or both.

Given this interest, it is desirable to identify additional tests that may probe the presence of dark matter cores in the general dwarf galaxy population. One possibility, which we explore in this contribution, is to use the susceptibility of dark matter haloes to tidal disruption, and its dependence on the inner density profile.

While cuspy NFW subhaloes appear to never fully disrupt when orbiting in smooth tidal fields (\citealt{Penarrubia2010,vdb2018,EP20}; \citealt{EN21}, hereafter \citetalias{EN21}), ``cored'' subhaloes should be more vulnerable to tides, and may eventually disrupt \citep{Penarrubia2010}. Cored subhaloes are particularly prone to tidal disruption in the inner regions of our Galaxy, where the Milky Way tidal field is strongest \citep{EPLG17}. The discovery of Milky Way satellites with small pericentric distances and short orbital periods, like the Tucana~3 dSph \citep[hereafer Tuc~3;][]{Drlica-Wagner2015,Shipp2018}, suggests that strong constraints on the presence of a core might be inferred from the apparent long-term survival of such objects.

We explore in this paper the tidal evolution of cored subhaloes, with the main goal of understanding under what conditions they may fully disrupt due to tides. This work extends our previous work on cuspy subhaloes (\citetalias{EN21}; \citealt{ENIP2022}), and explores a wide range of subhalo masses, core sizes, and orbital parameters. We make use of controlled $N$-body simulations where each subhalo is resolved by $10^7$ particles -- a resolution currently inaccessible to cosmological simulations. 

The paper is structured as follows: We begin by outlining our numerical set-up in Sec.~\ref{sec:num_setup}. Then, in Sec.~\ref{sec:TidalEvolution}, we study the systematics of the tidal evolution of cored subhaloes, and examine the relation between subhalo mass, core size and orbit as well as their relation to the criteria and time-scale for tidal disruption. We apply these results to the (surviving) satellites of the Milky Way in Sec.~\ref{sec:Application} and discuss the constraints they place on potential core sizes and the consistency of such constraints with current SIDM models. Finally, we summarize our main conclusions in Sec.~\ref{sec:Conclusions}.

\section{Numerical simulations}
\label{sec:num_setup}
We outline in this section the set-up of our $N$-body models, discussing in Sec.~\ref{sec:subhalo_models} the subhalo model, in Sec.~\ref{sec:host_and_galaxy} the host galaxy and orbits, and in Sec.~\ref{sec:nbody_code} the $N$-body code.

\begin{figure*}
  \begin{center}
  \includegraphics[width=\textwidth]{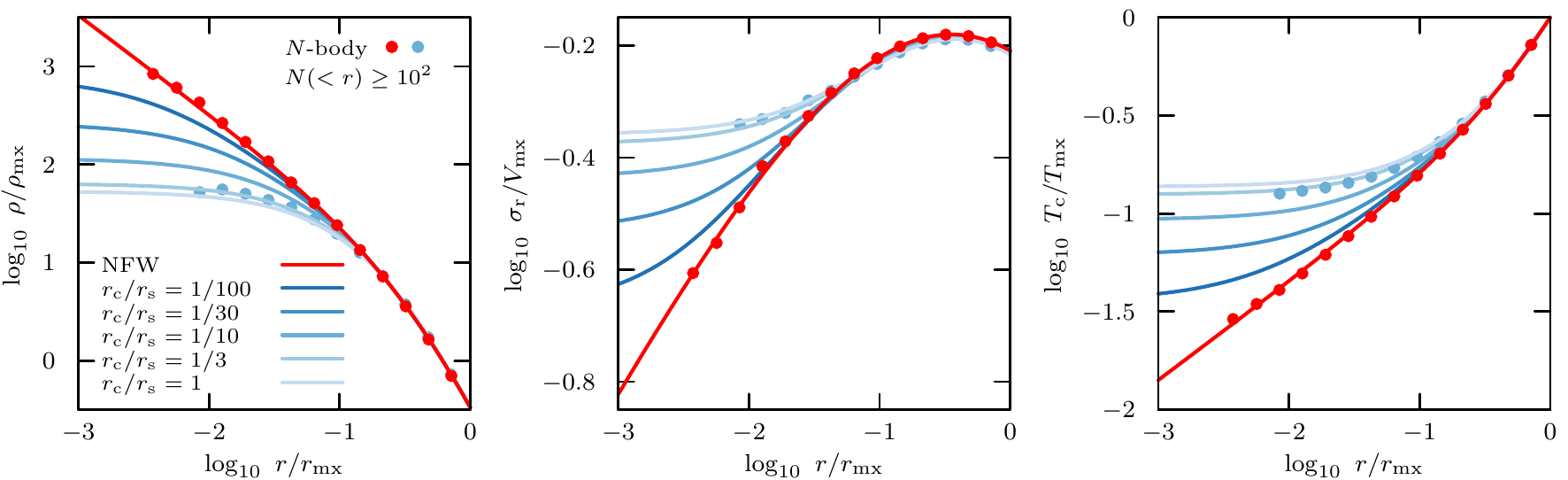}
  \end{center}
  \caption{The family of cuspy and cored subhalo density profiles given by Eq.~\ref{eq:cNFW}. The left-hand panel shows the density as a function of radius, normalized by $\rhomx = 3/(4 \pi)~ \Mmx \rmx^{-3}$ and $\rmx$, respectively. The NFW limit ($\rc=0$) is shown in red, while cored profiles for $1/100 \leq \rc/\rs \leq 1$ are shown in blue. The central panel shows the radial velocity dispersion $\sigma_\mathrm{r}$ (computed assuming isotropy), normalized by $\Vmx$. The cored models have a finite non-zero central velocity dispersion, whereas for the NFW limit, $\sigma_\mathrm{r}\rightarrow 0$ for $r \rightarrow 0$. The right-hand panel shows the periods of circular orbits $T_\mathrm{c}$ within the different subhaloes, normalized by $\Tmx$. For cored subhaloes, $T_\mathrm{c}$ flattens off towards the centre and converges to a finite non-zero value: orbits in the central core all have very similar orbital periods independent of radius. }
  \label{initial_profiles}
\end{figure*} 

\subsection{Subhalo models}
\label{sec:subhalo_models}
CDM subhaloes have been shown to follow a ``universal'' density profile whose shape is well described by the NFW formula, 
\begin{equation}
\label{eq:NFW}
 \rho_\mathrm{NFW}(r) =  \rho_\mathrm{s}  \left(r/r_\mathrm{s} \right)^{-1}   \left(1+r/r_\mathrm{s}\right)^{-2} ~. 
\end{equation}
The profile has two parameters, which may be chosen to be the scale density, $\rho_\mathrm{s}$, and the scale radius, $\rs$. The profile is ``cuspy'', with a central density that nominally diverges as
$\diff \ln \rho / \diff \ln r \rightarrow -1$ for $r \rightarrow 0$. 

In this work, we consider subhaloes that are NFW-like at radii larger than their scale radius $\rs$, but deviate from Eq.~\ref{eq:NFW} in the inner regions, where they exhibit a small central constant-density core, i.e., $\diff \ln \rho / \diff \ln r \rightarrow 0$ for $r \rightarrow 0$. We adopt the parametrisation of \citet{Penarrubia2012}, which adds a single ``core size'' parameter, $\rc$,  to the original NFW formula:
\begin{equation}
\label{eq:cNFW}
 \rho_\mathrm{cNFW}(r) =  \rho_\mathrm{s}  \left(r_\mathrm{c}/r_\mathrm{s} + r/r_\mathrm{s} \right)^{-1}  \left(1+r/r_\mathrm{s}\right)^{-2} ~. 
\end{equation}
This parametrisation is adequate to describe cores that are small relative to the scale radius $\rs$. For $\rc = 0$, Eq.~\ref{eq:cNFW} reduces to the original NFW formula. 
For $\rc > 0$, the central density converges to a finite value, 
\begin{equation}
 \rho_0 =   \rho_\mathrm{s}~ \rs / \rc ~.
\end{equation}
At the core radius\footnote{Core sizes of dark matter haloes are frequently expressed through their equivalent \citet{Burkert1995} core radius, $r_\mathrm{b}$, defined as the radius where the density drops to one quarter of the central value. To facilitate the comparison with prior work, we note that for the density profiles used in this work (Eq.~\ref{eq:cNFW}), the core radius parameter $r_\mathrm{c}$ is related to $r_\mathrm{b}$ through $r_\mathrm{c}/r_\mathrm{s}  =  r_\mathrm{b}/r_\mathrm{s}~[ 4(1+r_\mathrm{b}/r_\mathrm{s})^{-2} -1 ]^{-1}$, i.e., for $r_\mathrm{b} \ll \rs$, $r_\mathrm{c} \rightarrow r_\mathrm{b}/3$. All our models have $\rc/\rs \leq 1$, which translates to $r_\mathrm{b}/\rs \lesssim 0.59$.}, $\rc$, the density equals $\rho_\mathrm{cNFW}(\rc) = \rho_0 (1+\rc/\rs)^{-2}/2$. This implies that for $\rc/\rs \rightarrow 0$, the core radius approximates the distance where the density drops to about $1/2$ of the central value: $\rho_\mathrm{cNFW}(\rc) \rightarrow \rho_0/2$. We shall hereafter refer to haloes satisfying eq.~\ref{eq:cNFW} as ``cored NFW'' or ``cNFW'' haloes, for short.

The three parameters $\rho_\mathrm{s}$, $\rs$, and $\rc$ fully determine the density profile. Instead of using $\rho_\mathrm{s}$ and $\rs$ directly, we will characterize a subhalo using its maximum circular velocity, 
$\Vmx = \max\left[\Vc(r)\right] = \max\left[\sqrt{GM(<r)/r}\right]$, and the radius, $\rmx$, where this maximum is reached. 

The relation between $\rs$, $\rmx$, and core size $\rc$ is well approximated by the following relation:
\begin{equation}
 \rmx / \rs \approx 2.16  \left( 1 + 6.65\,  \rc/\rs \right)^{0.35},
\end{equation}
with an accuracy of better than one per cent for $0 < \rc/\rs < 1$. For $\rc = 0$, the NFW limit, $\rmx \approx 2.16\,\rs$.

The circular velocity corresponding to Eq.~\ref{eq:cNFW} has a closed analytical form \citep[][equation~3]{Penarrubia2012}. For convenience, we provide a fit for $\Vmx$,
\begin{equation}
  \Vmx / V_\mathrm{s} \approx 1.65  \left( 1 + 3.1\,  \rc/\rs \right)^{-0.20}~,
\end{equation}
where $V_\mathrm{s} = \rs \sqrt{ G \rho_\mathrm{s} }$. The fitting formula is accurate to better than $0.5$ per cent for $0 < \rc/\rs < 1$. For the NFW limit, $\Vmx \approx 1.65\, V_\mathrm{s}$.

For reference, we also define the mass enclosed within $\rmx$, 
\begin{equation}
 \Mmx = \rmx \Vmx^2 G^{-1}~,
\end{equation}
as well as the period of a circular orbit of radius $\rmx$,
\begin{equation}
 \Tmx = 2 \pi \rmx / \Vmx~.
\end{equation}
We shall refer to $\Vmx$, $\rmx$, $\Mmx$, and $\Tmx$ as the characteristic velocity, radius, mass, and time of a subhalo, respectively. The relation between $\Mmx$ and the virial mass, $M_{200}$, of the halo depends weakly on the halo concentration, defined as $c=r_{200}/\rs$. For the NFW limit $\rc/\rs=0$, for example, $\Mmx=0.31\,M_{200}$ and $\Mmx=0.22\,M_{200}$ for $c=10$ and $20$, respectively.

Density profiles using Eq.~\ref{eq:cNFW} are shown in the left-hand panel of Fig.~\ref{initial_profiles} for different choices of the core size, $\rc/\rs$. Radii are normalized to $\rmx$ and densities are normalized to $\rhomx = 3/(4\pi)~\Mmx \rmx^{-3}$. The NFW profile is shown in red, and cored profiles are shown in blue.

The central panel shows the radial velocity dispersion, $\sigma_\mathrm{r}$, for the same profiles (assuming isotropy). For NFW, $\sigma_\mathrm{r} \rightarrow 0$ for $r \rightarrow 0$, while the cored models have a finite and non-zero central velocity dispersion.

Finally, the circular orbit period, $T_\mathrm{c}(r) = 2 \pi r / \Vc(r)$, is shown as a function of radius in the right-hand panel of Fig.~\ref{initial_profiles}.
Note that for the NFW limit, $T_\mathrm{c} \rightarrow 0$ for $r \rightarrow 0$, while for $\rc > 0$, the orbital time has a well-defined central minimum,
\begin{equation}
 \label{eq:central_time}
 T_0 =  \lim_{r \rightarrow 0}~ 2 \pi r / \Vc(r) = \left( \frac{3 \pi}{G \rho_0} \right)^{1/2}=  \left( \frac{3 \pi \rc}{G \rho_s \rs} \right)^{1/2}.
\end{equation}

We generate $N$-body realizations of Eq.~\ref{eq:cNFW} with isotropic velocity dispersion by (i) computing the corresponding distribution function through Eddington inversion and by (ii) sampling from this distribution function. We make use of the implementation of \citet{EP20}, available online\footnote{\url{https://github.com/rerrani/nbopy}}. The density profiles are exponentially tapered beyond $10\,\rs$ to obtain numerical models with finite total mass. Each subhalo model is constructed with $10^7$ $N$-body particles, although we have also run selected simulations with $10^6$ particles to check for convergence. The convergence tests are presented in Appendix~\ref{Appendix:Convergence}.

\begin{table}
\centering
\caption{Overview of the initial $N$-body parameters. All subhalo models are spherical and have an isotropic velocity dispersion. We run 160 $N$-body models on a grid over
16 log-spaced values for the initial density contrast between subhalo and host as expressed through $T\mxzero/\Tperi$, five different core sizes $\rc/\rs$ and two orbits with peri-to-apocentre ratios $1\rt1$ and $1\rt5$.
All physical quantities, like, e.g., the subhalo mass $\Mmx$ or host circular velocity $V_\mathrm{host}$, are given for illustration only and may be re-scaled, keeping $V\mxzero \ll V_\mathrm{host}$. For convergence tests, the models with $\rc/\rs=0$ and $1/100$ on circular orbits were also run with a grid resolution of $r\mxzero/256$ (see table entries marked with a star).   }
\label{tab:simulation_overview}
\begin{tabular}{l@{\hskip 0.8cm} l @{\hskip 0.8cm}l}
\toprule
\multirow{8}{*}{\rotatebox[origin=t]{90}{\textsc{S u b h a l o e s}}}
 & Profile          & cNFW (Eq.~\ref{eq:cNFW}) \\
 & $\rc/\rs$        & {$0^\star$,~ $1/100^\star$,~ $1/30$,~ $1/10$,~ $1/3$}  \\
 & $T\mxzero/\Tperi$& {$0.3$,~ $\dots$,~ $4.4$} \\
 & $M\mxzero$       & {$10^6\, \Msol$}  \\[0.3cm]
 & $N$              & {$10^7$} \\ 
 & $\Delta x$       & {$r\mxzero/128$, $^\star r\mxzero/256$}\\
 & $\Delta t$       & {$\min ( T\mxzero,\Tperi )/400$} \\[0.3cm]
\multirow{4}{*}{\rotatebox[origin=t]{90}{\textsc{H o s t}}}
& Profile          & Isothermal (Eq.~\ref{eq:host_potential}) \\
& $V_\mathrm{host}$& {$220\,\kms$} \\
& $\rperi$         & {$40\,\kpc$} \\
& $\rperi\rt\rapo$ & {$1\rt1$,~ $1\rt5$ }\\

\bottomrule
\end{tabular}
\end{table}

\begin{figure*}
  \begin{center}
  \includegraphics[width=\textwidth]{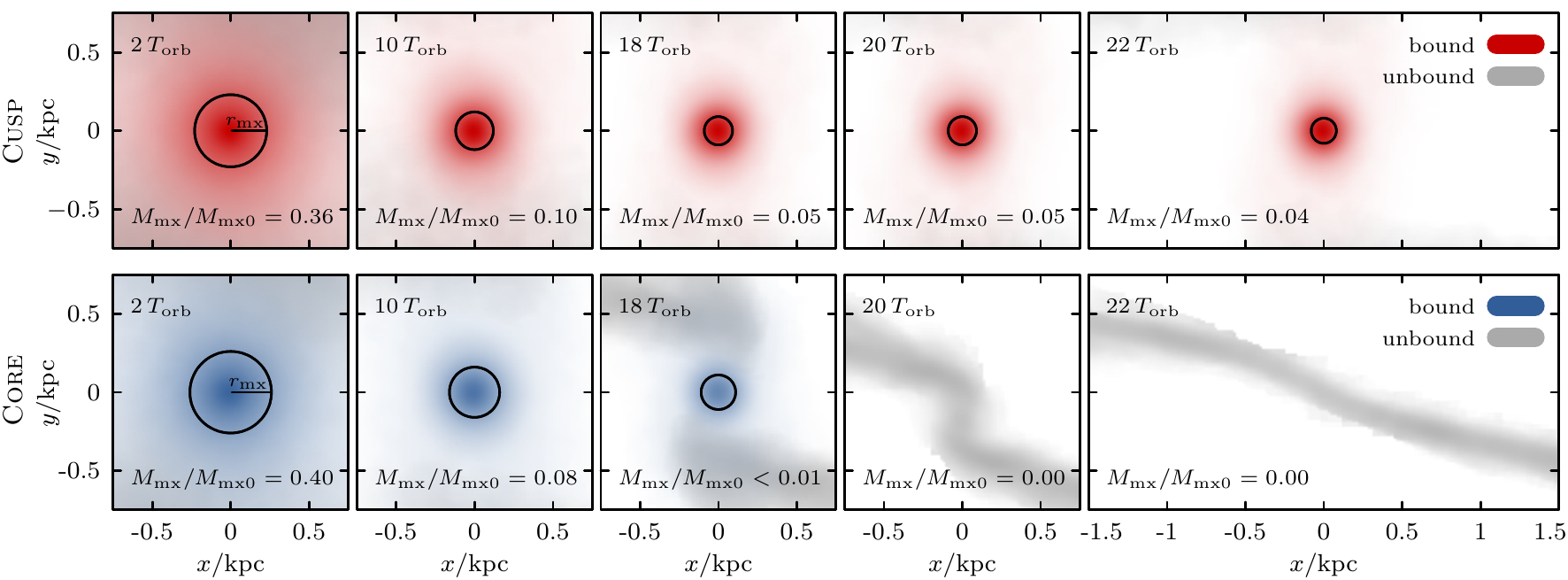}
  \end{center}
  \caption{Simulation snapshots of the tidal evolution of a cuspy (top row) and a cored (bottom row) dark matter subhalo on an orbit with a peri-to-apocentre ratio of $1\rt5$. Snapshots are taken at different apocentres after (from left to right) 2, 10, 18, 20 and 22 orbital periods $T_\mathrm{orb}$. Both subhaloes have the same initial characteristic mass, size and velocity. The cuspy subhalo is an NFW model, while the cored subhalo has a core size of $\rc/\rs=1/3$ (Eq.~\ref{eq:cNFW}). Particles bound to the respective subhalo are shown in blue, while unbound particles are shown in grey. A circle of radius $\rmx$, computed for the bound particles, is shown in black. While the cuspy subhalo converges relatively fast to a stable remnant state, the cored subhalo disrupts after 18 orbital periods. A video of this simulation is available on the journal website.}
  \label{fig:snapshots}
\end{figure*}

\subsection{Host galaxy model and orbits}
\label{sec:host_and_galaxy}

We evolve our $N$-body subhalo models in a spherical, static, isothermal host potential,
\begin{equation}
 \label{eq:host_potential}
 \Phi_\mathrm{host} = V_\mathrm{host}^2 ~ \ln \left( r / r_0 \right)~,
\end{equation}
with a constant circular velocity of $V_\mathrm{host}=220\, \kms$ ($r_0$ denotes an arbitrary reference radius). 
This choice of a scale free host model allows straightforward re-scaling of our simulation results, and provides a good approximation to the rather flat Milky Way circular velocity curve inferred between $5\,\kpc$ and $25\,\kpc$ (see Fig.~3 in \citealt{Eilers2019}).
The corresponding host virial mass and virial radius are $M_\mathrm{200} = 3.7\times10^{12}\,\Msol$, and $r_\mathrm{200} = 325\,\kpc$, respectively. 

Subhalo models are placed either on circular orbits or on eccentric orbits with a peri-to-apocentre ratio of $1\rt5$. The value of $1\rt5$ is close to the average peri-to-apocentre ratio of many Milky Way satellites \citep{Li2021a}.

In the potential of Eq.~\ref{eq:host_potential}, the period of a circular orbit with an orbital radius of $40\,\kpc$ equals $T_\mathrm{orb} = 1.1\,\Gyr$. On an eccentric $1\rt5$ orbit with $\rperi = 40\,\kpc$ and $\rapo = 200\,\kpc$, the radial period equals $2.5\,\Gyr$. Note that these values are given for illustration only, and, since the simulations are scale free, they may be re-scaled as needed to other physical values.

\subsection{N-body code}
\label{sec:nbody_code}
We use the particle mesh code \textsc{superbox} \citep{Fellhauer2000} to evolve our $N$-body models in the analytical host potential. The code uses two cubic grids that are moving with and are centred on the subhalo, as well as a static grid containing the full simulation volume. The linear resolution of the two co-moving high- and medium-resolution grids is $\Delta x \approx \rmx/128$ and $10\, \rmx/128$, respectively. The static grid has a lower resolution of $\approx 500\,\kpc/128$. Individual simulations have been repeated at higher grid resolution ($\Delta x \approx \rmx/256$) to test for numerical convergence of our results. 

Time integration is done using a leapfrog integrator with a constant time-step $\Delta t = \min(\Tperi, \Tmx)/400$. A circular orbit within an NFW subhalo at a radius equal to the grid resolution $\Delta x \approx \rmx/128$ is then resolved by $\approx 16$ time-steps. The same orbit in a cored subhalo with $\rc/\rs = 1/3$ is resolved by $\approx 52$ time-steps.

\begin{figure}
  \begin{center}
  \includegraphics[width=\columnwidth]{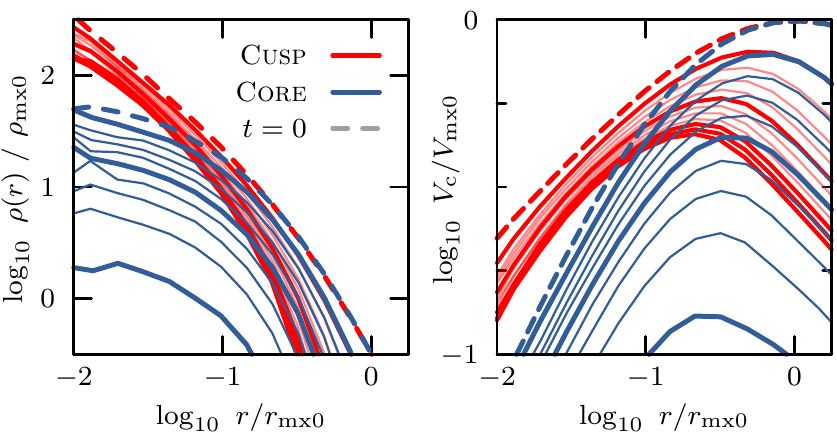}
  \end{center}
  \caption{Evolution of the density (left-hand panel) and circular velocity (right-hand panel) of the cuspy (NFW) and cored ($\rc/\rs=1/3$) subhalo models of Fig.~\ref{fig:snapshots}. Radii are normalized by the initial subhalo characteristic size $r\mxzero$, densities by $\rho\mxzero =3/(4\pi)~M\mxzero r\mxzero^{-3} $ and circular velocities by $V\mxzero$, which are identical for the cuspy and cored models. Each curve corresponds to a snapshot taken at apocentre ($t/\Torb=0,2,4,\dots$). Snapshots shown in Fig.~\ref{fig:snapshots} are highlighted using thick lines. Initial profiles are shown using dashed lines. While the cuspy subhalo converges relatively fast to a stable remnant state, the cored subhalo in this example continues to lose mass until, after 18 orbital periods, no bound particles remain.}
  \label{fig:density_evolution}
\end{figure}

\section{Tidal evolution}
\label{sec:TidalEvolution}

\begin{figure}
  \begin{center}
  \includegraphics[width=8.5cm]{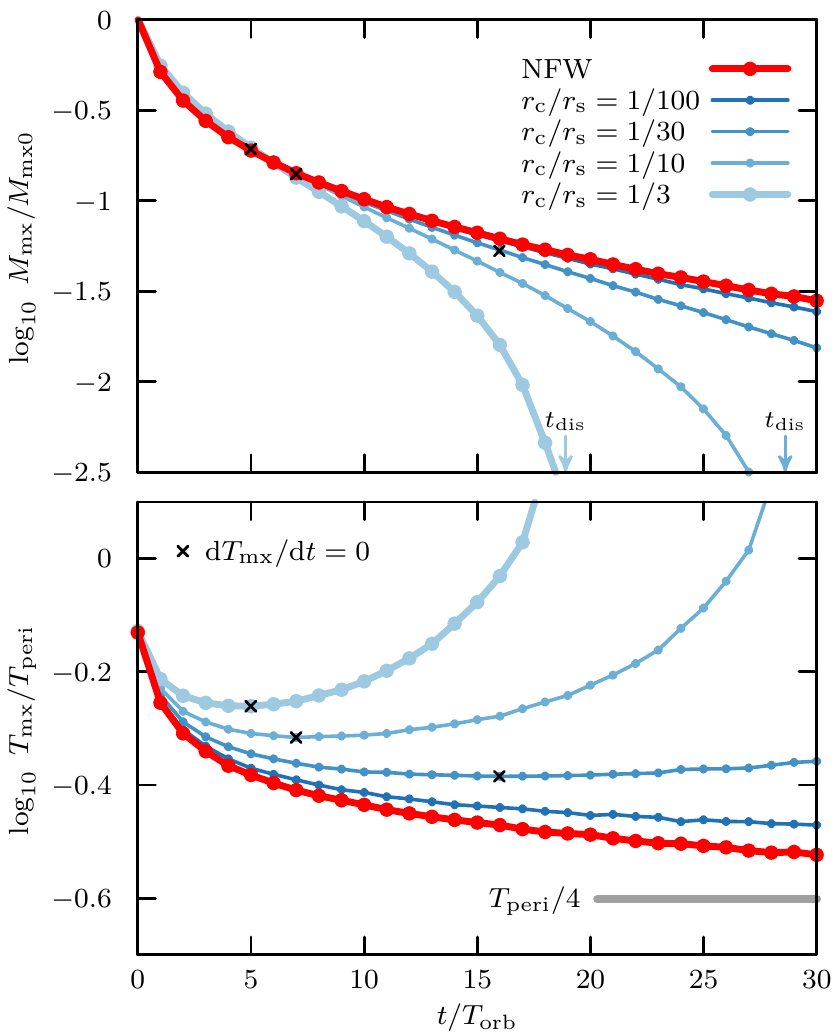}
  \end{center}
  \caption{Evolution of the characteristic mass (top panel) and time (bottom panel) of the bound remnants of subhaloes on $1\rt5$ eccentric orbits. All subhaloes have the same initial characteristic mass, size and velocity but different core radii. The models shown in Fig.~\ref{fig:snapshots} (NFW and $\rc/\rs=1/3$) are highlighted using thick curves. Cores make subhaloes more susceptible to tides, leading in some cases to full disruption. With increasing core size $\rc$, it takes fewer orbits to disrupt a subhalo. On the other hand, if the core size is sufficiently small, a stable bound remnant can be reached even in the presence of a core.}
  \label{fig:mmx_vs_time}
\end{figure}

\begin{figure}
  \begin{center}
  \includegraphics[width=8.5cm]{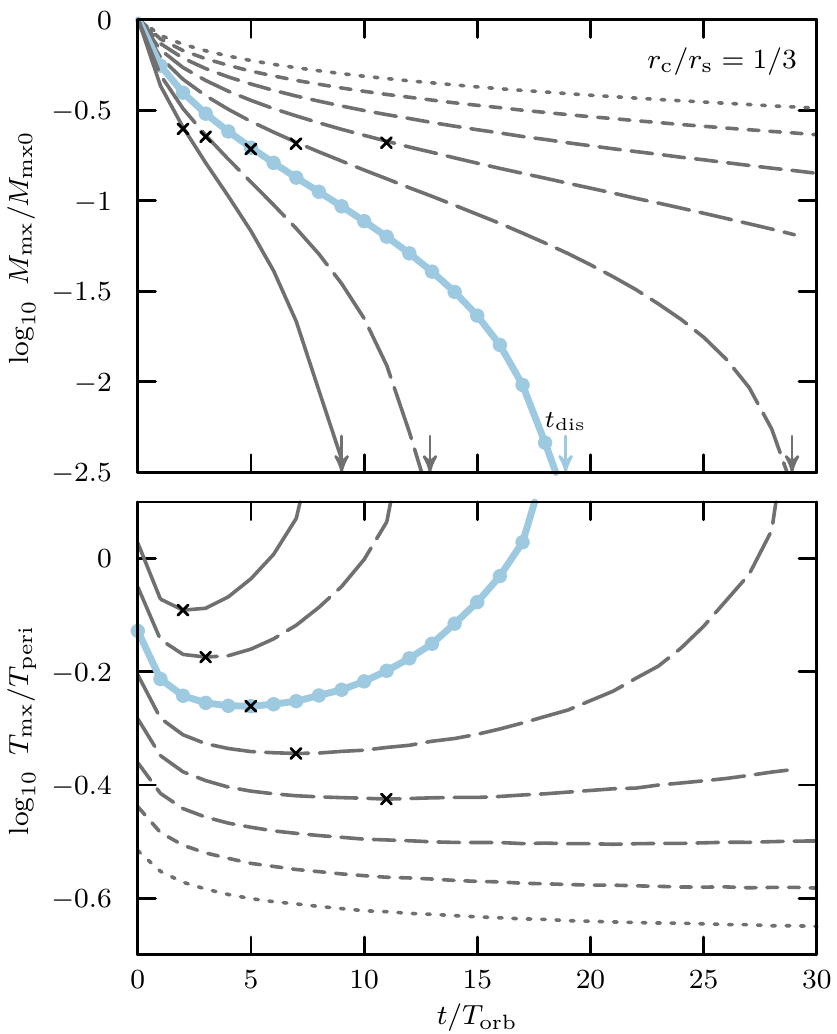}
  \end{center}
  \caption{Like Fig.~\ref{fig:mmx_vs_time}, for cNFW subhaloes with identical initial core size ($\rc/\rs=1/3$), but for eight different values of the initial density contrast between subhalo and the host halo at pericentre, expressed by $T\mxzero / \Tperi$.
  For subhaloes with small $T\mxzero / \Tperi$ (in this example, for $\log{T\mxzero / \Tperi} \lesssim -0.4$), the cNFW subhalo converges to a bound remnant. Subhaloes with larger initial $T\mxzero$ eventually fully disrupt. The larger $T\mxzero / \Tperi$, the shorter the time it takes till full disruption.} 
  \label{fig:mmx_vs_time_fix_profile}
\end{figure}

\subsection{Tidal evolution of cored and cuspy subhaloes}

We begin our analysis by comparing the tidal evolution of a cuspy (NFW) and a cored (cNFW) subhalo in the host potential described in Sec.~\ref{sec:host_and_galaxy}. The two subhaloes are chosen to have identical initial characteristic mass ($M\mxzero$), radius ($r\mxzero$), and velocity ($V\mxzero$). The cored subhalo in this example is chosen to have $\rc/\rs = 1/3$ (see eq.~\ref{eq:cNFW}).

The subhalo characteristic mass and velocity are chosen to be much smaller than those of the host in order to prevent effects like dynamical friction or orbital decay due to tides \citep{White1983, HernquistWeinberg1989, Fellhauer2007, Miller2020}, and to ensure that the orbital parameters of the bound remnant remain largely unchanged during evolution, simplifying the analysis.

For convenience, we scale the results to physical values corresponding to $M\mxzero =10^6\,\Msol$, $r\mxzero=0.42\,\kpc$, and $V\mxzero =3.2\,\kms$, respectively, but note that the results are scale free and may be scaled as needed to other values, keeping $V\mxzero \ll V_\mathrm{host}$. The two subhaloes are injected in the host potential at apocentre, and placed on an orbit with a pericentre $\rperi=40\,\kpc$ and an apocentre $\rapo = 200\,\kpc$. 

Figure~\ref{fig:snapshots} shows snapshots taken at different apocentric passages, after 2, 10, 18, 20 and 22 radial orbital periods $T_\mathrm{orb}$ (from left to right, respectively). 
The evolution of the subhalo density profiles and circular velocity curves are shown in Fig.~\ref{fig:density_evolution}. Dashed curves correspond to the initial profiles, and other curves are shown every two apocentric passages. Thick lines correspond to the snapshots highlighted in Fig.~\ref{fig:snapshots}.

As discussed in detail by \citetalias{EN21}, the NFW subhalo (top row) evolves quickly towards a well-defined stable bound\footnote{We compute self-bound particles by iteratively (i)~determining the subhalo centre through the shrinking spheres method \citep{Power2003}, (ii)~computing the particles' energies $E=v^2/2 + \Phi(r)$ under the assumption of spherical symmetry for the subhalo potential $\Phi(r)$, and (iii) removing those particles with positive total energy. The steps are repeated until convergence.} remnant. The cored subhalo (bottom row), on the other hand, disrupts fully\footnote{We define ``full disruption'' as the time, $t_\mathrm{dis}$, when the bound mass $\Mmx$ drops to less than $1/1000$ of its initial value and we are unable to resolve any remaining bound remnant. Since mass-loss accelerates before full disruption, our measure of $t_\mathrm{dis}$ is well-defined and insensitive to numerical resolution.} after $\sim18$ orbital periods, leaving behind no self-bound remnant.

This illustrates the main difference between the tidal evolution of cored and cuspy subhaloes: although NFW subhaloes are expected to always leave behind some stable bound remnant, cored subhaloes are subject to full disruption under certain conditions.

\subsection{Tidal disruption criteria for cored subhaloes}

Under what conditions do cored subhaloes disrupt fully, and how long does the disruption process take? We explore this by first varying the core size of the cNFW subhaloes discussed in the previous subsection, while keeping the orbit unchanged. The evolution of the characteristic mass and time of the bound remnants of these subhaloes is shown in Fig.~\ref{fig:mmx_vs_time} for $\rc/\rs=0$, $1/100$, $1/30$, $1/10$ and $1/3$. All of these systems have identical initial mean densities; i.e., equal values of $r\mxzero$ and $V\mxzero$.

As tides strip the NFW ($\rc/\rs=0$) subhalo, they leave behind a bound remnant with lower mass but higher characteristic density (i.e., shorter characteristic time) than the initial object. The NFW subhalo appears to converge to a stable remnant whose characteristic time approaches asymptotically $\Tperi/4$, where $\Tperi$ is the circular orbital time at pericentre. As discussed by \citetalias{EN21}, this is indeed the final fate of an NFW subhalo subject to heavy tidal stripping.

The cNFW subhalo with the smallest core in the series ($\rc/\rs=1/100$) follows a similar evolution to that of the NFW system, although the remnant seems to converge to a final density slightly lower (larger $\Tmx$) than the NFW remnant (see the bottom panel of Fig.~\ref{fig:mmx_vs_time}).

For larger core sizes, the evolution is qualitatively different; after initially decreasing, the characteristic time (density) of the remnant is seen to reach a minimum (maximum) and then gradually increase (decrease). This time (marked with small crosses in Fig.~\ref{fig:mmx_vs_time}) corresponds to a saddle point in the evolution of the bound mass beyond which the rate of mass-loss starts to accelerate and the characteristic density of the remnant starts to decrease (i.e., $\Tmx$ starts to increase). The bound remnant becomes gradually less and less dense, and it eventually fully disrupts\footnote{Similar to their NFW counterparts, the cored subhaloes studied here follow ``tidal evolutionary tracks'', i.e., mono-parametric functions that describe the evolution of subhalo structural parameters with respect to their initial conditions \citep{Penarrubia2008,Penarrubia2010}. We refer to Appendix~\ref{Appendix:TidalTracks} for a discussion of the tracks corresponding to cNFW density profiles.}.

The core size needed to avoid total disruption is likely to depend on the strength of the tidal perturbation experienced by the subhalo. 
To first order, the relative strength of the tides may be characterized by the density contrast between subhalo and host at pericentre, which may be quantified by the ratio between the initial characteristic subhalo time and the orbital period at pericentre, $T\mxzero/\Tperi$.

We therefore explore in Fig.~\ref{fig:mmx_vs_time_fix_profile} the evolution of subhaloes with given initial core size ($\rc/\rs=1/3$) on a fixed orbit ($\rperi=40\,\kpc$, $\rapo=200\,\kpc$, i.e., constant $\Tperi$), for a range of initial subhalo characteristic times $T\mxzero$. The model with $T\mxzero/\Tperi =0.7$, identical to the one shown in Fig.~\ref{fig:mmx_vs_time}, is shown in blue, while other orbits are shown in grey. For sufficiently small values of $T\mxzero/\Tperi$, the cored subhaloes seem to converge to a bound remnant and should survive indefinitely. On the other hand, cored subhaloes disrupt for large values of $T\mxzero/\Tperi$. The larger $T\mxzero/\Tperi$, the shorter the time to disruption.

\subsection{Disruption times}
\label{sec:DisruptionTimes}

\begin{figure}
  \begin{center}
  \includegraphics[width=8.5cm]{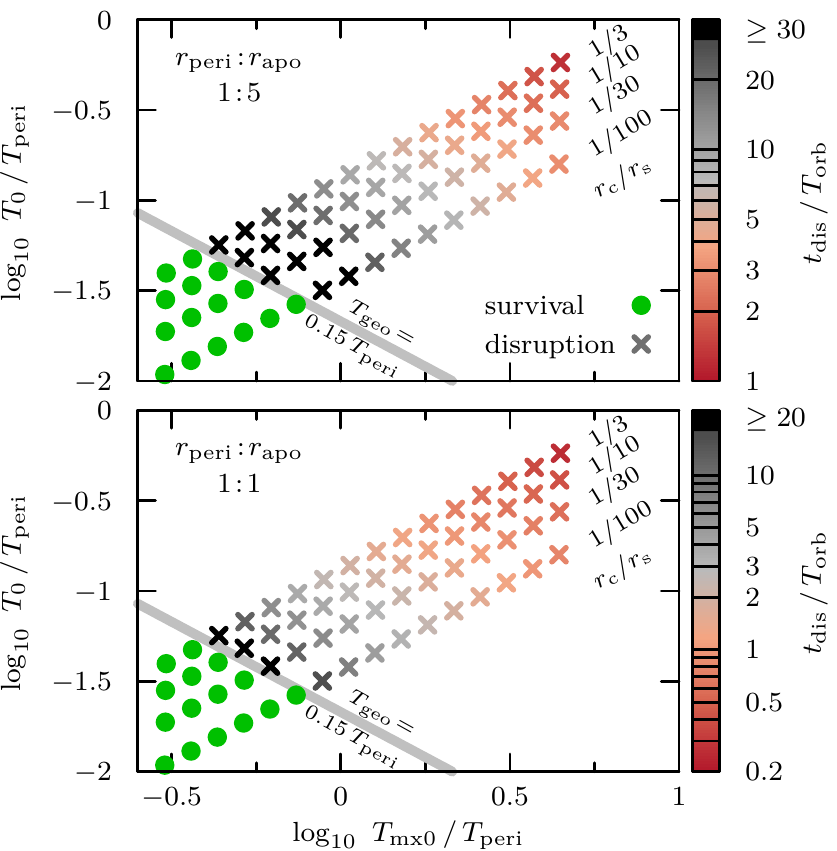}
  \end{center}
  \caption{Initial central times $T_0$ and characteristic times $T\mxzero$, normalized by the circular time at pericentre, $\Tperi$, of all simulation runs listed in Table~\ref{tab:simulation_overview}. Models of equal initial core size $\rc/\rs$ fall on the same diagonal line in this plane. Lines corresponding to different constant values of $\rc/\rs$ are parallel.
  Subhaloes on orbits with peri-to-apocentre ratio of $1\rt5$ are shown in the top panel, and those on circular orbits in the bottom panel. 
  Subhaloes that have fully disrupted within the simulated time (20 and 30 orbital periods for the circular and eccentric orbits, respectively), as well as those that at the end of the simulation have ``accelerating'' mass-loss ( $\diff^2 \Mmx / \diff t^2 < 0$ ) , are shown using crosses (``disruption'').
  The time of disruption (defined as the time when the bound remnant has less than $0.1\,$per cent of the initial mass, and can no longer be resolved) is colour coded.
  Subhaloes that at the end of the simulation have converged to a stable bound remnant or whose mass-loss is steadily decelerating ( $\diff^2 \Mmx / \diff t^2 > 0$ ) are shown using green circles (``survival''). 
  Neither the initial central time $T_0$ nor the initial characteristic time $T\mxzero$ are on their own sufficient to predict whether a subhalo will survive or disrupt.
  The line $ \Tgeo=(T_0 T\mxzero)^{1/2} \approx 0.15\,\Tperi$ (shown in grey in both panels) roughly separates the two regimes: i.e. for a given $\Tperi$, the geometric mean $\Tgeo\equiv (T_0 T\mxzero)^{1/2}$ may be used to predict the survival or eventual disruption of a subhalo. } 
  \label{fig:Tgeo_limit}
\end{figure} 

The results of the previous subsection suggest that the final fate of a cNFW subhalo depends on the interplay between the initial characteristic time(s) of the subhalo, and the orbital time at pericentre. Cored subhaloes are characterized by two different time-scales, one corresponding to its initial central density, $T_0$, and another one, $T\mxzero$, that describes its initial mean density. Our simulation suite, listed in Table~\ref{tab:simulation_overview}, covers a wide range of $T_0$ and $T\mxzero$, as well as a broad range of pericentric radii, or, equivalently, $\Tperi$.

Fig.~\ref{fig:Tgeo_limit} presents our full simulation grid (see Table~\ref{tab:simulation_overview}), after scaling $T_0$ and $T\mxzero$ to $\Tperi$. The top panel shows subhaloes evolved on eccentric orbits with peri-to-apocentre ratio of $1\rt5$, and the bottom panel shows the same models evolved on circular orbits. Each simulation is coloured by the value of $t_\mathrm{dis}$ (crosses are used for systems that disrupt in less than the total simulated time, or have accelerating mass-loss at the end of the simulation; circles otherwise). Crosses of similar colour denote systems with similar $t_\mathrm{dis}$, which evolve nearly indistinguishably from each other.

It is clear from Fig.~\ref{fig:Tgeo_limit} that, for given $\Tperi$, the fate of a cNFW subhalo is tied to a combination of both $T_0$ and $T\mxzero$. In particular, the product $T_0 T\mxzero/\Tperi^2$ seems to be the sole parameter needed to describe the disruption time (or survival) of a subhalo. In other words, for given $\Tperi$, a single characteristic time, 
\begin{equation}
\Tgeo\equiv\left(T_0\, T\mxzero\right)^{1/2}~,
\end{equation}
i.e., the geometric mean between $T_0$ and $T\mxzero$, seems to characterize fully the tidal evolution of a cNFW subhalo.

Do these results depend on the assumed eccentricity ($1\rt5$) of the simulated orbits? Our earlier work on NFW subhaloes \citepalias{EN21} suggests that the primary effect of the eccentricity is to delay the effect of tides on eccentric orbits relative to circular orbits with the same pericentre by some factor: $f_\mathrm{ecc}\approx 5$ for $1\rt5$ orbits, $\approx 6.5$ for $1\rt10$ and $\approx8$ for $1\rt20$. In other words, systems with equal values of $\Tgeo/\Tperi$ should require $f_{\rm ecc}\approx 5$ times more orbits till disruption when evolved on the $1\rt5$ eccentric orbits presented in Fig.~\ref{fig:Tgeo_limit} than on circular ones.

\begin{figure}
  \begin{center}
  \includegraphics[width=8.5cm]{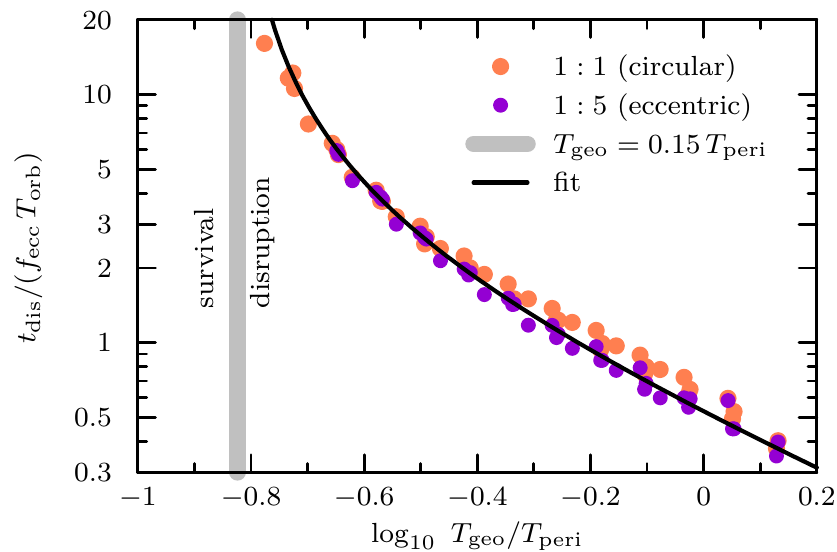}
  \end{center}
  \caption{Number of (circular) orbits till full disruption, $t_\mathrm{dis}/T_\mathrm{orb}$, as a function of the initial density contrast between subhalo and host, expressed through $\Tgeo/\Tperi$ (orange circles) . If the initial $\Tgeo/\Tperi$ is sufficiently small, an asymptotic remnant is reached and the subhalo does not disrupt. The threshold for the existence of an asymptotic remnant is shown as a vertical grey bar. Models with initial conditions that fall to the right of this bar disrupt eventually. The number of orbits till disruption decreases as $\Tgeo/\Tperi$ increases. Purple circles show the results for eccentric orbits (peri-to-apocentre ratio $1\rt5$), scaled by $f_{\rm ecc}\approx 5$ \citepalias{EN21}. The overlap between orange ($f_{\rm ecc}=1$) and purple ($f_{\rm ecc}\approx 5$) points suggests that the main effect of orbital eccentricity is to delay disruption compared to a circular orbit of equal pericentre by a factor $f_{\rm ecc}$. The fit of Eq.~\ref{eq:Ncirc_vs_Tgeo} is shown as a solid black curve.  }
  \label{fig:T_disrupt}
\end{figure}

This is apparently also the case for the cored models studied here, as shown in Fig.~\ref{fig:T_disrupt}. This figure shows the number of orbits needed to fully disrupt a system as a function of $\Tgeo/\Tperi$. Orange circles correspond to circular orbits ($f_{\rm ecc}=1$) and purple symbols to the $1\rt5$ eccentric orbits ($f_{\rm ecc}\approx 5$). Disruption times on eccentric orbits are clearly just delayed by roughly $f_{\rm ecc}$ relative to circular, as shown by the overlap between symbols of different colour.

In addition, in all cases the number of orbits needed for disruption increases with decreasing $\Tgeo/\Tperi$, and steepens as $\Tgeo/\Tperi$ values approach $0.15$. The following function (shown as a solid curve in Fig.~\ref{fig:T_disrupt}) reproduces the simulation results quite well:
\begin{equation}
\label{eq:Ncirc_vs_Tgeo}
 \frac{t_\mathrm{dis}}{f_\mathrm{ecc}~T_\mathrm{orb}} =   \begin{cases} 
                                                ~~3 ~  \displaystyle  \left( \frac{\Tgeo}{0.15\, \Tperi} - 1 \right)^{-1}  & \text{if }  \Tgeo >  0.15\, \Tperi\\
                                                ~~\infty & \text{otherwise. } 
                                               \end{cases}
\end{equation}

Note that $\Tgeo/\Tperi=0.15$ is the same boundary shown with a grey band in Fig.~\ref{fig:Tgeo_limit}, which neatly separates cNFW subhaloes that fully disrupt from those that survive for at least $30$ full orbits. We interpret this as implying that $\Tgeo \approx 0.15\, \Tperi$ is a simple but robust criterion determining the ultimate survival or disruption of a cNFW subhalo.

\section{Application to Milky Way satellites}
\label{sec:Application}

\begin{figure}
  \begin{center}
  \includegraphics[width=6.05cm]{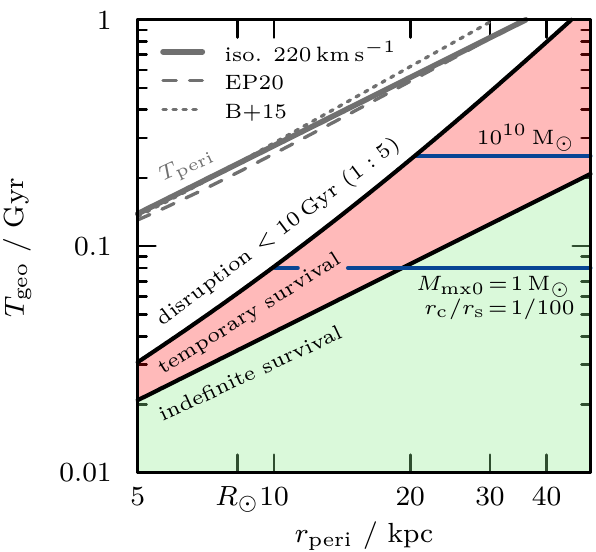}\includegraphics[width=2.45cm]{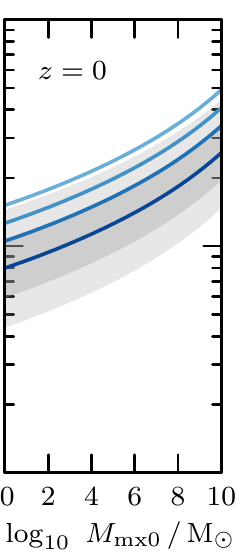}
  
  \includegraphics[width=8.5cm]{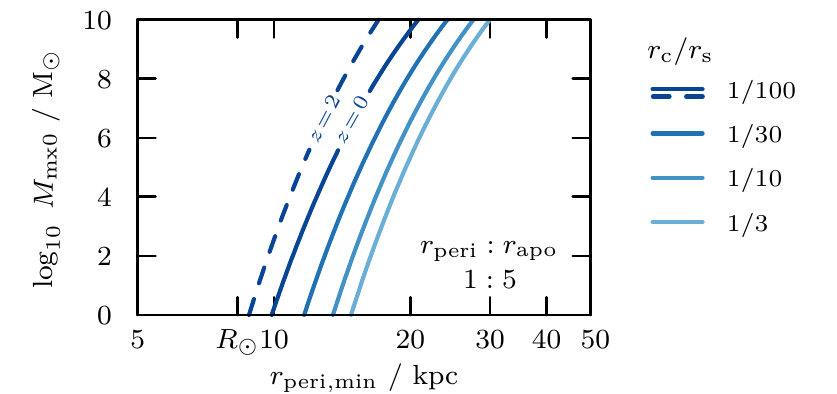}
  \end{center}
  \caption{{\it Top left:} Survival/disruption criteria for Milky Way subhaloes. The grey curves show $\Tperi$ as a function of pericentric radius $\rperi$ for various models of the inner regions of the Milky Way. The green shaded area indicates the values of $\Tgeo$ of subhaloes expected to survive indefinitely (``indefinite survival'').
  Subhaloes with an initial $\Tgeo$ in the pink-shaded area may survive for up to $10\,\Gyr$ on an orbit with a peri-to-apocentre ratio of $1\rt5$ (``temporary survival''). Subhaloes with even larger values of $\Tgeo$ disrupt in less than $10\,\Gyr$ (``disruption'', white area).
  For each initial value of $\Tgeo$, there is a well-defined minimum pericentric distance necessary for the temporary survival of a subhalo: two example values (for $\Tgeo=0.08\,\Gyr$ and $\Tgeo=0.25\,\Gyr$) are depicted as blue horizontal lines. 
  {\it Top right:} This panel shows $\Tgeo$ as a function of initial subhalo mass for cNFW subhaloes with four different core radii $\rc/\rs$, assuming that they follow the redshift $z=0$ LCDM mass-concentration relation (grey-shaded bands correspond to successive $\pm0.1\,\mathrm{dex}$ scatter in concentration).
  {\it Bottom:} This panel indicates, as a function of subhalo mass, the pericentric radii inside which subhaloes would disrupt in less than $10\,\Gyr$ on a $1\rt5$ orbit. At the solar circle, $R_\odot= 8\,\kpc$, virtually all subhaloes with masses $M\mxzero \gtrsim 1\,\Msol$ would have disrupted if they had core sizes larger than $\rc/\rs \gtrsim 1/100$. A dashed curve shows the same threshold derived using the $z=2$ mass-concentration relation.}
  \label{fig:minimum_rperi}
\end{figure}

The observed satellite population of the Milky Way may be compared with our results to derive upper limits on the core sizes of the subhaloes they inhabit (or, more precisely, on their $\Tgeo$ values) by assuming that they inhabit subhaloes that have survived disruption. We illustrate this in Fig.~\ref{fig:minimum_rperi}, where the grey diagonal curves in the top-left panel show the circular orbit time-scale, $\Tperi = 2\pi \rperi /V_\mathrm{host}$, as a function of pericentric radius for two different Milky Way mass models \citep{Bovy2015,EP20}, both of which are adequately approximated by the simple isothermal potential introduced in Sec.~\ref{sec:host_and_galaxy}.

To survive indefinitely, subhaloes must have $\Tgeo\lesssim0.15\, \Tperi$, which defines the green zone labelled ``indefinite survival'' in Fig.~\ref{fig:minimum_rperi}. Cored subhaloes with $\Tgeo$ above that zone will in principle disrupt on a time-scale given by eq.~\ref{eq:Ncirc_vs_Tgeo}. Assuming a $1\rt5$ orbital eccentricity, we can identify which cNFW subhaloes would disrupt in less than $10\,\Gyr$ (those in the zone coloured white). Those with initial $\Tgeo$ and $\Tperi$ in the pink zone labelled ``temporary survival'' would eventually disrupt but they may still have a self-bound remnant after $10\,\Gyr$ of evolution.

We compare these constraints with the $\Tgeo$ times expected for cNFW haloes with cores of different sizes in the top right-hand panel of Fig.~\ref{fig:minimum_rperi}. The four blue curves correspond to four different choices of $\rc/\rs$ spanning the range $1/100$ to $1/3$. The calculation assumes that cNFW haloes follow the same initial mass-concentration relation (or, equivalently, the same $r\mxzero$-$V\mxzero$ relation) as cuspy LCDM haloes, computed\footnote{We note that LCDM haloes have characteristic times, $\Tmx$, that depend only weakly on mass: $\Tmx \sim 1$ Gyr for $\Mmx=10^8\, M_\odot$, increasing (decreasing) by less than a factor of $\sim 1.4$ for $\Mmx=10^{10}\, M_\odot$ ($10^{6}\, M_\odot$), see fig.~13 in \citetalias{EN21}.} following \citet{Ludlow2016} for $z=0$. 

This panel shows that even for initial core sizes as small as $1\,$per cent of the scale radius (i.e., $\rc/\rs=1/100$), $\Tgeo$ is of the order of $\sim 0.25\,\Gyr$ for an $\Mmx \sim 10^{10}\, M_\odot$ subhalo, which implies that no such subhalo could survive on an orbit with $\rperi <20\,\kpc$ for $\sim 10$ Gyr. This case is indicated by a solid horizontal blue line in the top-left panel of Fig.~\ref{fig:minimum_rperi}, and corresponds to a core size of just $\rc\approx60\,\pc$.
This constraint becomes even more restrictive for larger core sizes. For $\rc/\rs=1/3$, or $\approx 2$ kpc for $\Mmx \sim 10^{10}\, M_\odot$, no such subhalo could survive for $10$ Gyr on orbits with $\rperi<30$ kpc.

Because of the weak dependence of $\Tmx$ on mass, not even cNFW subhaloes with masses as small as $1\, M_\odot$ could survive on orbits with $\rperi<10\,\kpc$. This is shown in the bottom panel of Fig.~\ref{fig:minimum_rperi} where we plot, as a function of subhalo mass, the minimum pericentre allowed if cNFW subhaloes are to survive for at least $10\,\Gyr$ in the MW tidal field. As in other panels, different curves correspond to different values of $\rc/\rs$. The dashed curve labelled ``$z=2$'' shows how much the $\rc/\rs=1/100$ curve shifts when using the redshift $z=2$ LCDM mass-concentration relation. Subhaloes accreted early by the Milky Way would be slightly denser, increasing their chance of survival, but the effect on the minimum pericentre is rather small.

We conclude that any mechanism that may impose a core as small as $\rc/\rs \gtrsim 1/100$ on LCDM haloes would lead to a remarkably smooth inner Milky Way halo, with virtually no dark matter substructures more massive than $1\,\Msol$ surviving till the present day \citep[see also][]{Penarrubia2010}. The presence of satellites with small pericentric radii and relatively short orbital times thus places particularly strong constraints on any potential core radius. We examine next the particular case of Tucana~3, a satellite with unusually small pericentric distance, to quantify better these constraints.

\subsection{The case of Tucana 3}
\label{sec:Tuc3}
Tuc~3 is a low-luminosity satellite of the Milky Way ($L \sim 10^4\,\mathrm{L}_\odot$, \citealt{Drlica-Wagner2015}) with an associated stellar tidal stream spanning $\sim 5^\circ$ on the sky \citep{LiTucana2018, Shipp2018}. The combined measurements of velocity dispersion ($ \sigma_\mathrm{los} = 0.1^{+0.7}_{-0.1}\,\kms$; \citealt{Simon2017}) and half-light radius ($44\pm6\,\pc$; \citealt{Drlica-Wagner2015}) suggest an upper limit on the mass-to-light ratio within the half-light radius of $M/L \lesssim 240$ \citep{Simon2017}. These data are consistent with Tuc~3 being a dark matter-dominated dwarf spheroidal (dSph) galaxy, but it does not exclude the possibility that it may have been an unusually large globular cluster before disruption. On the other hand, Tuc~3 has stars that are unusually rich in r-process elements \citep{Hansen2017,Marshall2019}, and such stars are not known to exist in globular clusters. The argument below assumes that Tuc~3 is a dark matter-dominated dSph in the process of being tidally disrupted.

Tuc~3 has the smallest pericentre of all known Milky Way ultrafaints. Taking into account the effect of the LMC on Tuc~3's orbit, \citet{Erkal2018} report a remarkably small pericentric distance of $\sim 3.5\,\kpc$, and an apocentre of $\rapo \sim 56\,\kpc$. The same authors argue that to reproduce its stream length, Tuc~3 must have completed at least three pericentric passages. Although it is difficult to completely exclude the possibility that Tuc~3 has only completed one orbit on the basis of its tail morphology, the short orbital time ($\sim 0.65$ Gyr) also favours the completion of multiple orbits.

\begin{figure}
  \begin{center}  
  \includegraphics[width=6.05cm]{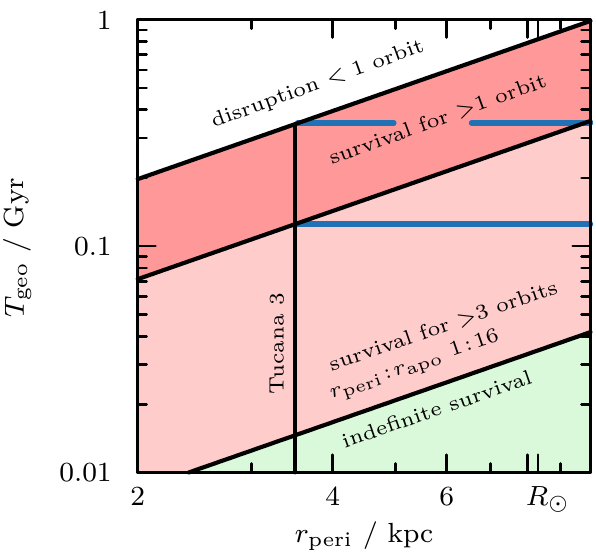}\includegraphics[width=2.45cm]{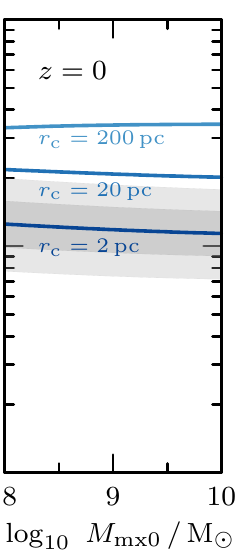}
  
  \end{center}
  \caption{As Fig.~\ref{fig:minimum_rperi}, but for a peri-to-apocentre ratio of $1\rt16$ ($f_{\rm ecc}\approx7.6$; see \citetalias{EN21}), chosen to approximate the nearly radial orbit of the Tuc~3 dwarf galaxy. Subhaloes with initial $\Tgeo$ in the green-shaded region survive indefinitely, while those in the red and pink-shaded regions survive for more than one or three orbits, respectively. At the pericentric distance of Tuc~3 ($\rperi\approx3.5\,\kpc$), this requires a core size of less than $\sim 200\,\mathrm{pc}$ for an $M\mxzero \sim 10^9\,\Msol$ subhalo, assuming that Tuc~3 has passed through pericentre at least once. If Tucana~3 has been on the current orbit for three orbital periods (unlikely given its well-developed tidal tails), the maximum allowed core size shrinks to $\sim 2\,\mathrm{pc}$.    }
  \label{fig:Tuc3}
\end{figure}

We can use these properties to derive an upper limit to the size of a potential core in Tuc~3. The argument follows closely our prior discussion of
Fig.~\ref{fig:minimum_rperi}. Indeed, the left-hand panel of Fig.~\ref{fig:Tuc3} is equivalent to Fig.~\ref{fig:minimum_rperi}, but extended to smaller pericentric radii, shorter values of $\Tgeo$, and orbits with peri-to-apocentre ratio of $1\rt16$. As in Fig.~\ref{fig:minimum_rperi}, subhaloes initially in the green-shaded region have core radii small enough to survive indefinitely. To survive more than three orbits, $\Tgeo\lesssim0.13\,\Gyr$, which corresponds to a core radius not greater than $\sim 2\,\pc$, assuming an initial subhalo mass in the range of $10^8$-$10^{10}\, M_\odot$. The constraint may be relaxed if Tuc~3 is completing its first pericentric passage (unlikely as that might be), but even in that case the core size should not exceed $200\,\pc$, or roughly 3~per~cent of $\rs$ at $\Mmx = 10^{10}\, M_\odot$. These are rather strict constraints, which may be compared with the core sizes expected, for example, in self-interacting dark matter models (SIDM), an issue we address in Section~\ref{subsec:SIDM}.

\subsection{Overview of constraints for Milky Way dwarf galaxies}

\begin{figure}
  \begin{center}
  \includegraphics[width=8.5cm]{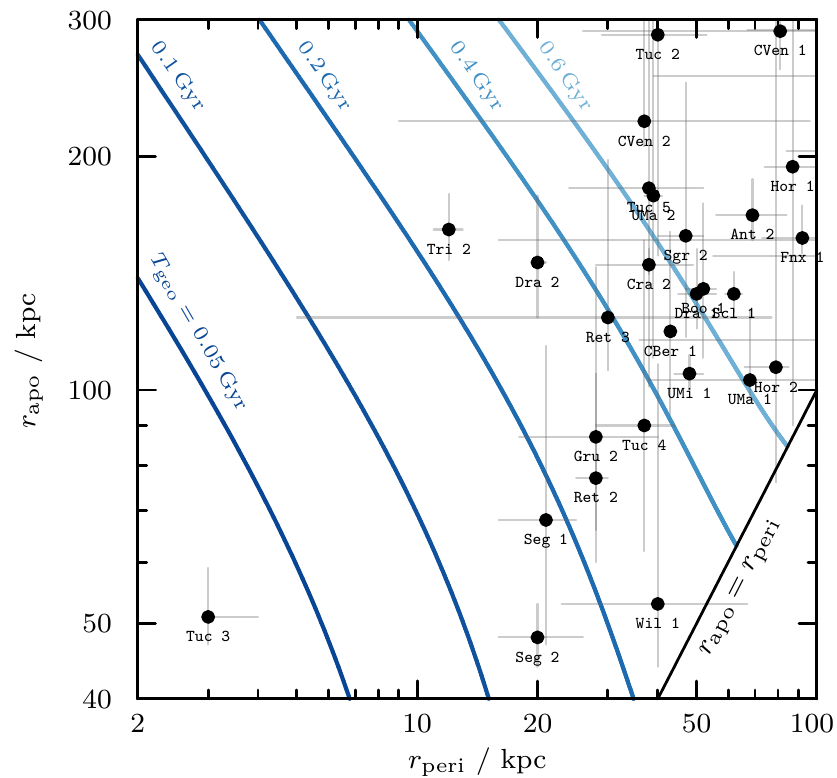}
  \end{center}
  \caption{Constraints on the tidal survival and disruption of cNFW subhaloes with different initial values of $\Tgeo$, assuming an evolution of $10\,\Gyr$ in an isothermal potential with $V_\mathrm{host}=220\,\kms$. For given $\Tgeo$, the $\rperi$ - $\rapo$ plane splits into two regions: one of survival, to the right of each line, and one of disruption, to the left. Filled black circles with errorbars show $\rperi$ versus $\rapo$ of Milky Way satellites, as computed by \citet{Li2021a} using Gaia EDR3 data. The orbit of Tuc~3 is strikingly different from those of other known Milky Way dwarfs, and places strong upper limits on its $\Tgeo$. The orbits of Seg~1 and Seg~2 constrain their respective values of $\Tgeo$ to be $ \lesssim 0.2\,\Gyr$, which corresponds to a core size of $\rc \lesssim 20\,\pc$, assuming a subhalo mass of $M\mxzero = 10^8 \sim 10^{10}\,\Msol$ (see the right-hand panel of Fig.~\ref{fig:Tuc3}). }
  \label{fig:MW_rperi_rapo_plane}
\end{figure}

We now extend the discussion on the tidal survival and disruption of cored subhaloes to a wider sample of Milky Way satellite galaxies. For this, we make use of the orbital parameters derived by \citet{Li2021a} using Gaia EDR3. The \citet{Li2021a} sample consists of 46 Milky Way dwarf galaxies in total. Here, we use the orbits of galaxies for which at least four member stars with spectroscopic measurements are included in the EDR3 catalogue. Further, we include only systems with pericentric distances $\rperi \leq 100\,\kpc$ and apocentres $\rapo \leq 300\,\kpc$ (25 in total).

As shown in Fig.~\ref{fig:T_disrupt} (see also Eq.~\ref{eq:Ncirc_vs_Tgeo}), disruption times, $t_\mathrm{dis}$, are proportional to the orbital time, with a proportionality constant that depends on orbital eccentricity and on the ratio $\Tgeo/\Tperi$. This implies that the condition $t_\mathrm{dis}=10\,\Gyr$ may be used to identify which subhaloes would be able to survive (or disrupt), depending on their apocentric and pericentric distances. 

The blue curves in Figure~\ref{fig:MW_rperi_rapo_plane} show the result of using Eq.~\ref{eq:Ncirc_vs_Tgeo} to compute which subhaloes should have disrupted/survived over $10\,\Gyr$ of evolution for different values of $\Tgeo$. Each curve divides the $\rperi$ versus $\rapo$ plane into two regions: one of disruption, to the left of the curve, and one of survival, for at least $10\,\Gyr$, to the right of it.

The particular nature of the Tuc~3 orbit is striking, requiring $\Tgeo \ll 0.05\,\Gyr$ (for a survival time of $10\,\Gyr$). The ultra faint dwarfs Seg~1 and Seg~2 require $\Tgeo \lesssim 0.2\,\Gyr$ to survive, which constrains their core sizes to $r_\mathrm{c}\lesssim20\,\pc$ (see right-hand panel of Fig.~\ref{fig:Tuc3}). For classical Milky Way satellites, like the Fornax dSph, the constraints are weaker: the limiting values of $\Tgeo$ exceed those of cNFW haloes with typical LCDM concentrations even for cores as large as $\rc/\rs\sim1$ (see right-hand panel of Fig.~\ref{fig:minimum_rperi}). 

We use next the results of Fig.~\ref{fig:MW_rperi_rapo_plane} to discuss the tidal survival of Milky satellites in the context of SIDM.

\subsection{SIDM-induced cores}
\label{subsec:SIDM}

Constant-density cores in NFW-like haloes may arise as a consequence of dark matter self-interactions (i.e., ``collisions''), which may ``heat up'' the inner cusp, reducing the central densities and leading to core formation \citep{Burkert2000, Spergel2000, Colin2002, Vogelsberger2012, Rocha2013, Kahlhoefer2019, Sameie2020}. The size of such cores in individual haloes is closely related to the self-interaction cross-section, which is usually assumed to lie in the range $0.1< s_\mathrm{si}/(\cm^2\g^{-1})<1$ on galaxy scales, where $s_\mathrm{si}=\sigma_\mathrm{SIDM}/m$ is the specific self-interaction cross-section \citep[for a review, see][]{Tulin2018}. Values in this range lead to tangible changes in the inner density profiles of dark matter haloes that may help to explain the observational results in apparent tension with cuspy dark matter haloes discussed in Sec.~\ref{SecIntro}.

The formation of cores through self-interactions should make subhaloes more vulnerable to tidal effects. Typically, larger self-interactions lead to larger cores, although the trend may reverse for extreme values of the cross-section because of the possibility of promoting the ``core collapse'' of the inner regions. The studies of \citet{Elbert2015} and \citet{Zeng2022} suggest that, for dwarf galaxy haloes, this only occurs for (velocity-independent) cross-sections exceeding $10\,\cm^2 \g^{-1}$, outside the range we consider here. We note, however, that this conclusion has recently been challenged by \citet{Nishikawa2020}, who have argued that tidal mass-losses may actually facilitate core collapse for velocity-dependent interaction cross-sections. 

Subhaloes in SIDM are also vulnerable to ``evaporation'' through the collision-mediated heat transfer from the hot host halo to the cooler substructures as subhaloes orbit the host. Indeed, this was one of the main arguments that led to skepticism about the viability of SIDM after it was proposed by \citet{Spergel2000}. The analytical arguments of \citet{GnedinOstriker2001} placed strong constraints on the allowed values of the cross-section, although such constraints were later relaxed by the results of direct N-body studies, which reported that subhalo evaporation was much weaker than analytical estimates \citep{Rocha2013}.

We focus below on the constraints that the tidal disruption of subhaloes described in earlier sections places on subhalo core sizes, without considering the effects of ``evaporation''. Neglecting subhalo evaporation makes any constraint on SIDM cross-sections we are able to place rather conservative, as its inclusion could only help dissolve substructure faster.

\subsubsection{Cross-section dependence of $\Tgeo$ in SIDM}

The tidal survival of subhaloes depends on the value of $\Tgeo$ imposed by the formation of a core. The mass and cross-section dependence of $\Tgeo$ may be estimated through direct numerical simulation of the collisional effects introduced by self-interactions. A number of cosmological SIDM simulations have been published in the recent past, although few have targeted the dwarf galaxy halo regime. We analyse here the published results of \citet{Rocha2013}, \citet{Elbert2015}, and a resimulation of one of the APOSTLE volumes (AP01-L1 in the notation of \citealt{Fattahi2016}) using the SIDM modifications to the EAGLE code described in \citet{Robertson2017, Robertson2018}.

\begin{figure}
  \begin{center}
  \includegraphics[width=8.5cm]{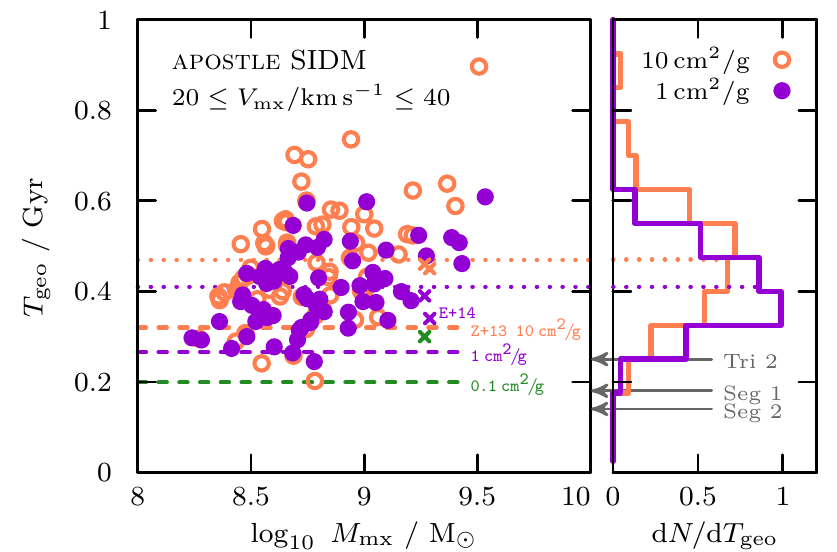}
  \end{center}
  \caption{Characteristic times, $\Tgeo=(T_0 \Tmx)^{1/2}$, for haloes in the APOSTLE SIDM simulations, as a function of halo mass, $\Mmx$. 
  The haloes are selected to be \emph{isolated} (i.e., not subhaloes) and to have circular velocities in the range $20 \leq \Vmx / \kms \leq 40$ (assumed to be the typical circular velocities of the subhaloes hosting Milky Way dwarf satellites, \citealt{Fattahi2018}).
  Values are measured for simulations with two different self-interaction cross-sections: $10\,\cm^2\g^{-1}$ (orange, open circles) and $1\,\cm^2\g^{-1}$ (purple, filled circles). 
  The medians are shown as dotted lines.
  In addition to the APOSTLE data, we show values of $\Tgeo$ measured from the \citet{Elbert2015} haloes for $10$, $1$ and $0.1\,\cm^2\g^{-1}$ (orange, purple and green crosses, respectively). The median values of $\Tgeo$ for the \citet{Zavala2013} \emph{subhaloes} are also shown using dashed lines. 
  For reference, upper limits on $T_\mathrm{geo}$ as constrained by MW satellites (see Fig.~\ref{fig:MW_rperi_rapo_plane}) are shown in the right-hand panel. } 
  \label{fig:SIDM_Tgeo}
\end{figure}

We measure $T_0$ and $\Tmx$ (and, hence, $\Tgeo$) for {\it isolated} haloes (i.e. ``centrals''; excluding subhaloes of larger systems), approximating the central densities by the mean densities within the innermost $300\,\pc$. Fig.~\ref{fig:SIDM_Tgeo} shows the estimated values of $\Tgeo$ as a function of halo mass for $s_\mathrm{si}=1\,\cm^2\g^{-1}$ (SIDM$_1$, shown using purple filled circles), and for $s_\mathrm{si}=10\,\cm^2\g^{-1}$ (SIDM$_{10}$, shown using orange open circles). These results are consistent with those estimated from the published density profiles of \citet{Elbert2015}, who follow the formation of two SIDM$_{1}$ and two SIDM$_{10}$ dwarf galaxy haloes (orange and purple crosses in Fig.~\ref{fig:SIDM_Tgeo}). 
Smaller cross-sections are expected to lead to smaller cores, and, consequently, smaller values of $\Tgeo$ at given mass. Indeed, the green crosses in Fig.~\ref{fig:SIDM_Tgeo} show the values of $\Tgeo$ measured for the dwarf galaxy halo of \citet{Elbert2015} for $s_\mathrm{si}=0.1\,\cm^2\g^{-1}$ (SIDM$_{0.1}$).

In addition to the above estimates obtained from individual \emph{isolated} haloes, we show values of $\Tgeo$ computed from the median density profiles of \emph{subhaloes} of the simulations of \citet{Zavala2013} for SIDM$_{0.1}$, SIDM$_{1}$ and SIDM$_{10}$, shown using dashed lines in green, purple and orange, respectively. Note that these values of $\Tgeo$ for subhaloes are slightly lower than those of isolated haloes. 
This is expected for tidally stripped systems, as $\Tmx$ decreases in the early stages of tidal evolution (see Fig.~\ref{fig:mmx_vs_time} and \ref{fig:mmx_vs_time_fix_profile}). 

\subsubsection{SIDM cores and the survival of Milky Way satellites}

We can now compare the $\Tgeo$ values shown Fig.~\ref{fig:SIDM_Tgeo} to the constraints illustrated in Fig.~\ref{fig:MW_rperi_rapo_plane}. For SIDM$_1$, for example, subhaloes should have $\Tgeo$ in the range of $0.3$ to $0.5\,\Gyr$. This seems consistent with most Milky Way satellites, except for a group of satellites with small pericentres and short orbital times (i.e., relatively small apocentres) that should have disrupted in less than 10 Gyr. The group includes Seg~1, Seg~2, Ret~2, Tri~2, and Wil~1, as well as the Tuc~3 dSph discussed earlier in Sec.~\ref{sec:Tuc3}.

Reducing the cross-section results in smaller core radii and enhances the ``survival'' region. However, even for SIDM$_{0.1}$, $\Tgeo$ is still of the order of $0.2 \sim 0.3 \,\Gyr$, and there is still a group of satellites in apparent conflict with SIDM-induced cores.

Before concluding that this conflict means that SIDM with cross-sections in the range of $0.1$--$1\,\cm^2\g^{-1}$ may be rejected, it is important to list some important caveats and qualifications.
\begin{itemize}
 \item[(i)] Dark matter may be self-interacting with a cross-section smaller than $0.1\,\cm^2\g^{-1}$. This would improve the survival of subhaloes on small pericentric orbits, but at the cost of making SIDM-induced cores much smaller, hindering the ability of SIDM to explain, for example, the slowly rising rotation curves of some dwarfs. 

 \item[(ii)] Satellite galaxies apparently in conflict with SIDM may have been accreted on to the Milky Way much more recently than $10\,\Gyr$, and therefore be transient systems in the process of being disrupted. This seems unlikely given the short orbital times of these objects, which suggest an early accretion into the Milky Way halo.

 \item[(iii)] Satellite galaxies apparently in conflict with SIDM may correspond to subhaloes made more resilient to tides after having undergone ``core collapse'' \citep{Colin2002,Kaplinghat2019,Zavala2019,Sameie2020}. Initial studies suggested the time-scales for core collapse to likely exceed the Hubble time, at least for velocity-independent cross-sections \citep{Balberg2002}. The more recent simulations by \citet{Zeng2022} confirm this and show that evaporation, triggered by the interaction of subhalo and host halo particles, may further delay core collapse. However, for models with velocity-dependent cross-sections, \citet{Nishikawa2020} argue that tides may shorten the time-scale for core collapse significantly. 

 \item[(iv)] The observed satellites might be the high-density tail of an otherwise disrupted population. Application of our criteria for disruption to tidal streams without known progenitor may allow to infer properties of this disrupted population. We plan to explore this in a future contribution.  
\end{itemize}
Fully cosmological SIDM $N$-body simulations of the tidal evolution of dwarf subhaloes on small pericentric orbits are likely needed to settle conclusively these questions.

\section{Summary and Conclusions}
\label{sec:Conclusions}

We have used N-body simulations to study the tidal evolution of NFW-like dark matter subhaloes with constant-density cores. These cored subhaloes differ from cuspy NFW profiles because of a finite and non-zero central velocity dispersion, and a well-defined minimum in the orbital time-scale at the centre. We evolve numerical realizations of cored subhaloes orbiting a host potential modelled as an isothermal sphere with a circular velocity $V_\mathrm{host}=220\,\kms$. Our suite of simulations explores a broad range of possible core sizes, orbital eccentricities, and pericentric radii. Our main findings may be summarized as follows:

\begin{itemize}
 \item[(i)] In agreement with previous studies, we find that, although well-resolved cuspy NFW haloes apparently always leave a bound remnant, cored subhaloes may fully disrupt in smooth tidal fields. 

 \item[(ii)] In the case of cuspy subhaloes heavily affected by tides, mass-loss gradually slows down as the remnant approaches a characteristic density determined by the host mean density at pericentre. On the other hand, as cored haloes get stripped beyond a certain point, their mass-loss rate accelerates, leading to full disruption. 
 
 \item[(iii)] Disruption times depend strongly on core size and on the orbital pericentre, and they approach infinity for small enough core radii. We find a simple criterion to identify which cored subhaloes survive indefinitely and which will fully disrupt. Cored subhaloes survive indefinitely if their inner characteristic crossing time satisfies $\Tgeo=(T_0 \Tmx)^{1/2} \lesssim 0.15\, \Tperi$, and they eventually disrupt otherwise. We provide a simple fitting formula (Eq.~\ref{eq:Ncirc_vs_Tgeo}) that expresses disruption times solely in terms of $\Tgeo$ (the geometric mean between the initial central orbital time, $T_0$, and the initial characteristic time, $T\mxzero=2\pi r\mxzero/V\mxzero$, of a subhalo) and $\Tperi$, the circular orbital time at pericentre.
 
 \item[(iv)] We apply these findings to subhaloes with structural parameters consistent with LCDM, and find that, even for core radii as small as $1$ per cent of their NFW scale radius, virtually all Milky Way subhaloes with masses $\Mmx \gtrsim 1\,\Msol$ would have fully disrupted in the inner $\sim10\,\kpc$ of the Milky Way. This means that the central halo of the Milky Way would be very smooth for dark matter models that predict core sizes larger than 1~per cent of the NFW scale radius. 
 
 \item[(v)] Applying our results to the Tuc~3 dSph constrains its core size to be no larger than $\sim2\,\pc$ if Tuc~3 has passed trough pericentre at least three times (as suggested by the clear presence of tidal tails), and to less than $\sim200\,\pc$ if Tuc~3 has passed through pericentre only once.  
 
 \item[(vi)] Applied to the core sizes expected for self-interacting dark matter (SIDM) models with specific velocity-independent cross-sections in the range of $0.1< s_\mathrm{si}/(\mathrm{cm}^2\mathrm{g}^{-1})<1$, we find that they are inconsistent with the long-term survival of a number of ultrafaint Milky Way satellites with small pericentric radii. These satellites have therefore either been accreted very recently (unlikely given their short orbital times), have core collapsed to account for their survival, or represent the high-density tail of an otherwise disrupted population. 
\end{itemize}

Although definitive conclusions regarding the viability of SIDM models need to await fully collisional cosmological N-body simulations of the evolution of substructure in a Milky Way-like halo, the tight constraints on core sizes placed by the survival of Milky Way dwarfs favour a simple interpretation where the subhaloes hosting ultrafaint Milky Way satellites are cuspy, as expected in the LCDM cosmological paradigm.

\footnotesize{
\subsection*{Acknowledgements}
The authors would like to thank Andrew Robertson, Aaron Ludlow and Isabel Santos-Santos for the access to density profiles of APOSTLE SIDM field haloes. We further thank Jens St\"ucker, Raul Angulo and Denis Erkal for insightful discussions. 
BF, RE and RI acknowledge funding from the European Research Council (ERC) under the European Unions Horizon 2020 research and innovation programme (grant agreement number 834148). 

\subsection*{Data availability}
The data underlying this article will be shared on reasonable request to the corresponding author.

\bibliography{cores}
}

\appendix

\section{Convergence tests}
\label{Appendix:Convergence}

\begin{figure}
  \begin{center}
  \includegraphics[width=8.5cm]{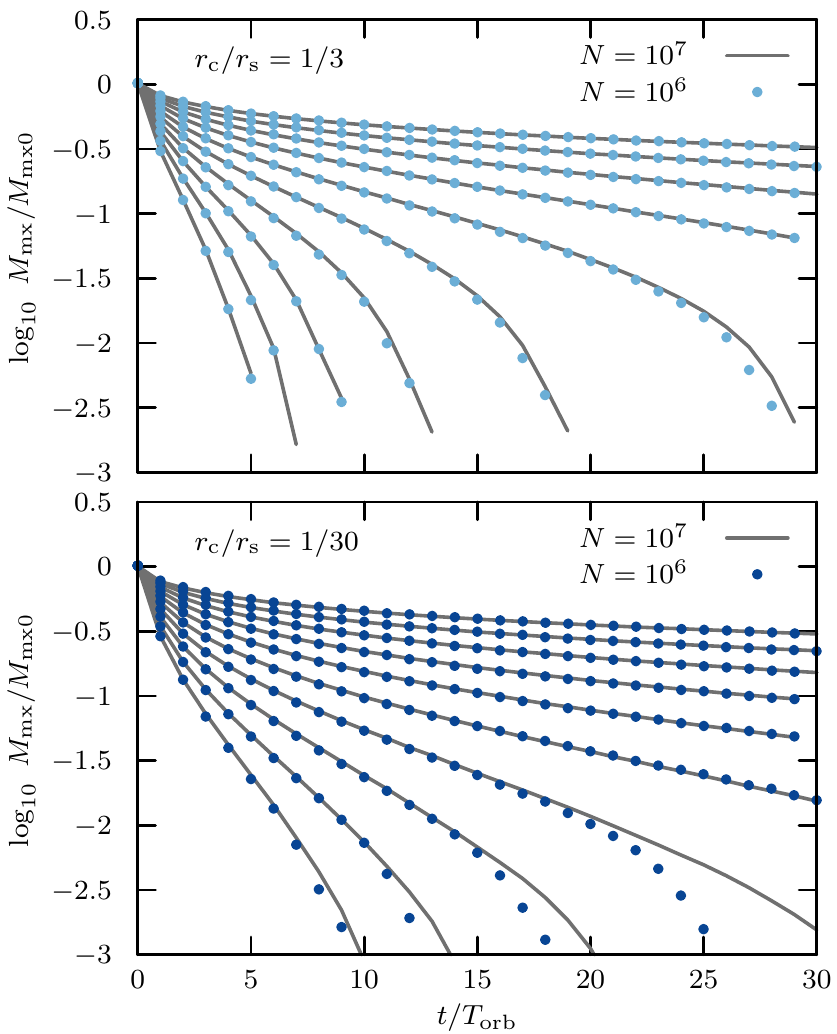}
  \end{center}
  \caption{Evolution of the bound mass fraction $\Mmx / M\mxzero$ as a function of time $t/\Torb$ for subhaloes on an orbit with peri-to-apocentre ratio $1\rt5$. \emph{Top panel: } mass evolution of ten subhaloes with a fixed core size of $\rc/\rs=1/3$ and different initial density contrasts corresponding to $0.3 \leq T\mxzero/\Tperi  \leq 1.5$. Models run with $N=10^7$ particles are shown as grey lines, and those with $N=10^6$ as blue filled circles. Only snapshots taken at apocentres are shown. \emph{Bottom panel:} like the top panel, but for subhaloes with an initial core size of $\rc/\rs = 1/30$. The mass evolution between the $N=10^6$ and $N=10^7$ runs is virtually indistinguishable for most models; the largest difference in disruption times (amounting to less than 20 per cent) occurs for models that take more than 20 orbits to disrupt.}
  \label{fig:convergence}
\end{figure}

To test how numerical resolution affects our results, we compare simulation runs with different numbers of $N$-body particles. For the NFW models discussed in this work, we use the same numerical set-up as in \citetalias{EN21}, and refer to appendix~A in \citetalias{EN21} for detailed convergence tests of those models. 
For the cored models, we repeat selected simulations of our simulation grid (Table~\ref{tab:simulation_overview}) on orbits with a peri-to-apocentre ratio of $1\rt5$ with $N=10^7$ and $N=10^6$ particles, for initial density contrasts between subhalo and host corresponding to $0.3 \leq T\mxzero/\Tperi  \leq 1.5$. 
The mass evolution of these models, for core sizes of $\rc/\rs=1/3$ and $\rc/\rs = 1/30$, is shown in Figure \ref{fig:convergence}. Simulations with $N=10^7$ are shown as grey curves, while those with $N=10^6$ are shown as blue filled circles. 
For most models, the mass evolution is virtually identical between the runs with $10^6$ and $10^7$ particles. Deviations are visible for models that take $\gtrsim20$ orbits to be stripped to $\Mmx/M\mxzero \leq 1/1000$, though the difference in disruption times remains less than $\sim 20$ per cent. 

\newpage
$~$

\newpage
\section{Tidal evolutionary tracks}
\label{Appendix:TidalTracks}

\begin{figure}
  \begin{center}
  \includegraphics[width=8.5cm]{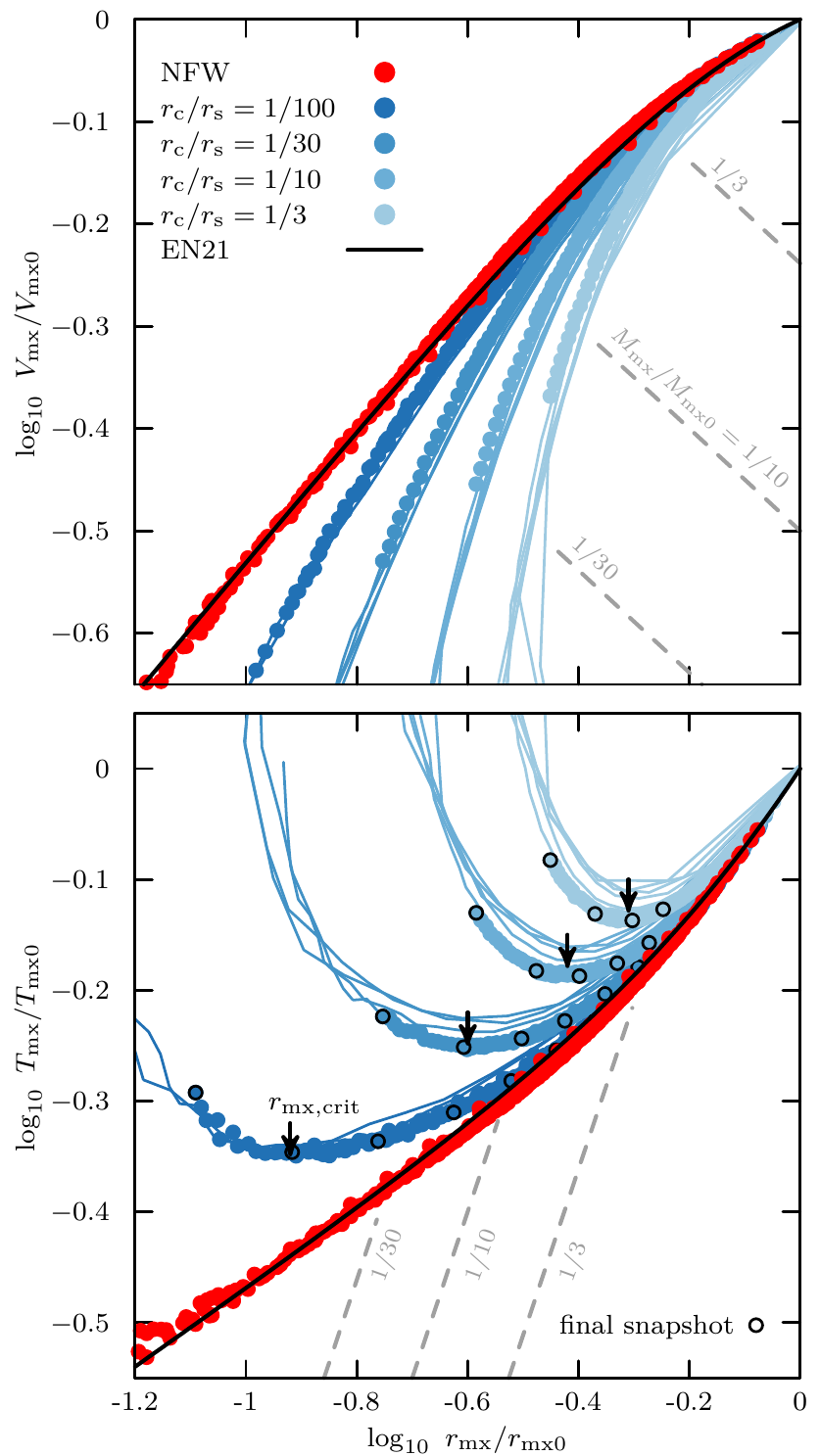}
  \end{center}
  \caption{Tidal evolutionary tracks of NFW (red) and cored subhaloes (blue) of the models listed in Table~\ref{tab:simulation_overview} (with $0.3 \leq T\mxzero/\Tperi  \leq 1.5$, on an orbit with peri-to-apocentre ratio of $1\rt5$). \emph{Top panel:} evolution of the characteristic velocity $\Vmx$ as a function of the characteristic size $\rmx$, normalized to their initial values. For reference, the tidal track for NFW haloes is shown as a black curve \citepalias{EN21}. Cored models have a lower $\Vmx / V\mxzero$ at fixed $\rmx / r\mxzero$ than their NFW counterparts. \emph{Bottom panel:} evolution of characteristic time $\Tmx / T\mxzero$. While for NFW models the characteristic time monotonously decreases with mass-loss, for cored systems, $\Tmx / T\mxzero$ increases once the system has been stripped beyond some threshold value $r_\mathrm{mx,crit}/r\mxzero$ (black arrows).}
  \label{fig:cored_tracks}
\end{figure}

Similar to their cuspy counterparts, cored subhaloes follow tidal evolutionary tracks \citep{Penarrubia2008,Penarrubia2010}, i.e., mono-parametric functions that describe how the subhalo structural parameters change during tidal mass-loss with respect to their initial values.
The top panel of Fig.~\ref{fig:cored_tracks} shows the evolution of the subhalo characteristic velocity $\Vmx/ V\mxzero$ as a function of its characteristic size $\rmx / r\mxzero$.
The evolution is shown for subhaloes with different initial core sizes (blue curves). Cuspy (NFW) models are shown in red. The structural properties are measured from snapshots taken at apocentres and are shown for the simulations on $1\rt5$ orbits listed in Table.~\ref{tab:simulation_overview} with $0.3 \leq T\mxzero/\Tperi  \leq 1.5$. Cored models that fully disrupt within the simulated time span of $30$ orbits are shown using lines, while those models where a bound remnant is resolved at the end of the simulation are shown using filled circles, each circle corresponding to an apocentre snapshot. For reference, we also show the NFW track of \citetalias{EN21} as a solid black curve. Tidal evolution progresses ``from right to left'' in these plots.

In the early stages of tidal evolution, the tracks of the cored subhaloes coincide with the NFW model. The larger the core size, the sooner the cored tracks start to deviate from the NFW track. With increasing core size, the tracks deviate systematically towards lower $\Vmx/V\mxzero$ at fixed $\rmx/r\mxzero$. 

The bottom panel of Fig.~\ref{fig:cored_tracks} shows the evolution of the subhalo characteristic time, normalized to its initial value.
For the NFW model, $\Tmx$ monotonously decreases during tidal stripping. The cored models, initially, follow this trend. However, after some tidal stripping, $\Tmx$ starts to increase, triggering a runaway disruption. The larger the core size $\rc/\rs$, the sooner a critical size $r_\mathrm{mx,crit}/r\mxzero$ is reached beyond which $\Tmx$ increases. This implies that for each core size $\rc/\rs$ there is a minimum characteristic size  $r_\mathrm{mx,crit}/r\mxzero$ down to which a subhalo may be stripped before disruption becomes inevitable. 
The existence of stable remnants for cored subhaloes therefore hinges on their tidal evolution slowing down sufficiently before surpassing that critical point.
The final snapshots of those simulations where a bound remnant is resolved at the end of the simulation are marked using black open circles. Only those subhaloes whose evolution stalls before reaching the critical point will never fully disrupt. Numerically, we find that subhaloes with core size parameters $\rc/\rs = 1/100$, $1/30$, $1/10$ and $1/3$ that are stripped to a size smaller than $r_\mathrm{mx,crit}/r\mxzero \approx$ 10, 25, 40 and 50 per cent, respectively, will eventually fully disrupt. These critical sizes are equivalent to minimum bound mass fractions $M_\mathrm{mx,crit}/M\mxzero$ of approx. 1, 5, 15 and 20 per cent, respectively.

\label{lastpage}

\end{document}